\documentclass{sig-alternate}
\makeatletter
\def\@copyrightspace{\relax}
\makeatother
\usepackage{caption}
\usepackage{tabularx}
\usepackage{subfigure}
\usepackage{multirow}
\usepackage{moresize}
\usepackage{morefloats}
 \usepackage{amsmath}
 \usepackage{indentfirst}
\usepackage{parskip}
\usepackage[ruled,vlined]{algorithm2e}
\usepackage{balance}
\begin{document}

\title{Spider and the Flies : Focused Crawling on Tumblr to Detect Hate Promoting Communities}
\author{
\numberofauthors{2} 
\alignauthor
Swati Agarwal\\
 \affaddr{Indraprastha Institute of Information Technology, Delhi (IIIT-D), India}\\
 \email{swatia@iiitd.ac.in}
\alignauthor
Ashish Sureka\\
 \affaddr{ABB Corporate Research Labs}\\
 \affaddr{Bangalore, India}\\
 \email{ashish.sureka@in.abb.com}
}
\maketitle
\begin{abstract}
Tumblr is one of the largest and most popular microblogging website on the Internet. Studies shows that due to high reachability among viewers, low publication barriers and social networking connectivity, microblogging websites are being misused as a platform to post hateful speech and recruiting new members by existing extremist groups. Manual identification of such posts and communities is overwhelmingly impractical due to large amount of posts and blogs being published every day. We propose a topic based web crawler primarily consisting of multiple phases: training a text classifier model consisting examples of only hate promoting users, extracting posts of an unknown tumblr micro-blogger, classifying hate promoting bloggers based on their activity feeds, crawling through the external links to other bloggers and performing a social network analysis on connected extremist bloggers. To investigate the effectiveness of our approach, we conduct experiments on large real world dataset. Experimental results reveals that the proposed approach is an effective method and has an F-score of $0.80$. We apply social network analysis based techniques and identify influential and core bloggers in a community.
\end{abstract}


\section{Introduction}
Tumblr is the second largest microblogging platform, has gained phenomenal momentum recently. It is widely used by fandoms: communities of users having similar interests in various TV shows and movies \cite{renwick2014audience}. Therefore, it is especially popular among young generation users and provides them a platform to discuss daily events. They communicate by blogging and publishing GIF images as their reactions and emotions on several topics \cite{Bourlai:2014:MCT:2615569.2615697}. According to Tumblr statistics $2015$ \footnote{https://www.tumblr.com/about}, over $219$ million blogs are registered on Tumblr and $420$ million are the active users. $80$ million posts are being published everyday, while the number of new blogs and subscriptions are $0.1$ million and $45$ thousands respectively.\\
\indent Tumblr is also posed as a social networking website that facilitates users to easily connect to each other by following other users and blogs without having a mutual confirmation. Bloggers can also communicate via direct messages that can be sent privately or can be posted publicly using 'ask box'. It facilitates bloggers to send these messages anonymously if they don't want to reveal their Tumblr identity \cite{marquart2010microblog}. Similar to other social networking websites, Tumblr has very low publication barriers. A blogger can publish a new post and can re-blog an existing public post which is automatically broadcasted to it's followers unless it is enabled as a private post \cite{DBLP:journals/corr/ChangTIL14}. The type of posts can be chosen among seven different categories including multi-media and other content: text, quote, link, photo, audio, video and URL. Unlike Twitter, Tumblr has no limit on the length of textual posts. Similar to hashtags in Twitter, there are separated tags associated with the blog content that make a post easier to be searchable on Tumblr \footnote{https://www.tumblr.com/docs/en/using\_tags}. Tumblr also allows users to update their other connected social networking profiles when something is posted.
\begin{figure*}[ht!]
\centering{
\includegraphics[width=0.65\textwidth]{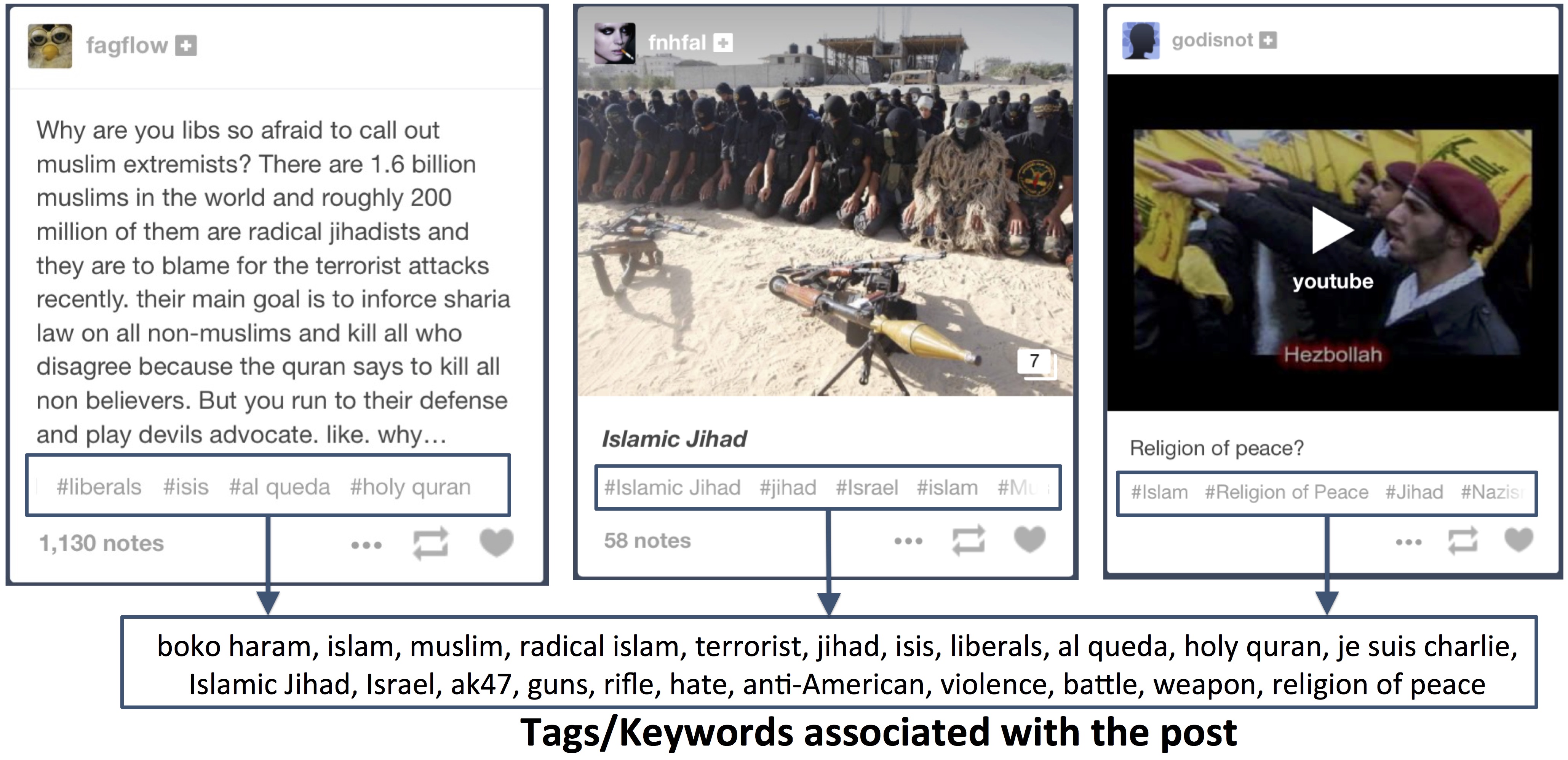}}
\caption{\label{tumblr_post_example}A Concrete Example of Different Types (Text, Video, Audio) of Hate Promoting Post on Tumblr}
\end{figure*} 

\indent The simplicity of navigation, high reachability across wide range of viewers, low publication barriers, social networking and anonymity has led users to misuse Tumblr in several ways. Previous studies shows that these features of Tumblr are exclusive factors to gain the attention of modern extremist groups \cite{bates2014psychological}\cite{sureka2014learning}. This is because Tumblr provides every kind of multimedia posts which is a great medium to share your views with your target audiences. These groups form their own communities that share a common propaganda. They post rude comments against a religion to express their hatred and spread extremist content. Social networking facilitates these groups to recruit more people to promote their beliefs and ideology among global audiences \cite{decary2011gang}\cite{mahmood2012online}. Figure \ref{tumblr_post_example} illustrates a concrete example of various types of hate promoting posts and their associated tags on Tumblr. The number of notes shows the number of times that post has been liked and re-blogged by other blogs.\\
\indent Online-radicalization and posting hateful speech is a crime against the humanity and mainstream morality; it has a major impact on society \cite{decary2011gang}. Presence of such extremist content on social media is a concern for law enforcement and intelligence agencies to stop such promotion in country as it poses the threat to the society \cite{Agarwal2015}\cite{devore2012exploring}. It also degrades the reputation of the website and therefore is a concern for website moderators to identify and remove such communities. Due to the dynamic nature of website, automatic identification of extremist posts and bloggers is a technically challenging problem \cite{wilner2011transformative}. Tumblr is a large repository of text, pictures and other multimedia content which makes it impractical to search for every hate promoting post using keyword based flagging. The textual posts are user generated data that contain noisy content such as spelling, grammatical mistakes, presence of internet slangs and abbreviations. Presence of low quality content in contextual metadata poses technical challenges to text mining and linguistic analysis \cite{Agarwal2015}\cite{marquart2010microblog}. The work presented in this paper is motivated by the need of investigating solutions to counter and combat the online extremism on Tumblr.\\
\indent The research aim of the work presented in this paper is the following:
\begin{enumerate}
\item To investigate the application of topical crawling based algorithm for retrieving hate promoting bloggers on Tumblr. Our aim is to examine the effectiveness of random walk in social network graph graph traversal and measuring its performance.
\item To investigate the effectiveness of contextual metadata such as content of the body, tags and caption or title of a post for computing the similarity between nodes in graph traversal. To examine the effectiveness of \textit{re-blogging} and \textit{like} on a post as the links between two bloggers. 
\item To conduct experiments on large real world dataset and demonstrate the effectiveness of proposed approach in order to locate virtual and hidden communities of hate and extremism promoting bloggers and apply Social Network Analysis based techniques to locate central and influential users.
\end{enumerate}
\section{Literature Survey}
In this section, we discuss closely related work to the study presented in this paper. Based on our review of existing work, we observe that most of the researches for detecting online radicalization are performed on Twitter, YouTube and various discussion forums \cite{agarwal2015applying}. We conduct a literature survey in the area of identifying hate promoting communities on social networking websites and short text classification of Tumblr microblog . O'Callaghan et. al. \cite{o2013uncovering} describe an approach to identify extreme right communities on multiple social networking websites. They use Twitter as a possible gateway to locate these communities in a wider network and track dynamic communities. They perform a case study using two different datasets to investigate English and German language communities. They implement a heterogeneous network within a homogeneous network and use four different social networking platforms (Twitter accounts, Facebook profiles, YouTube channels and all other websites) as extreme right entities or peers and edges are the possible interactions among these accounts.\\
\indent Mahmood S. \cite{mahmood2012online} describes several mechanisms that can be useful in order to detect presence of terrorists on social networking websites by analyzing their activity feeds. They use Google search and monitor terror attack using keyword-based flagging mechanism. They monitor sentiments and opinions of users following several terrorism groups on on-line social networks and propose a counter-terrorism mechanism to identify those users who are more likely to commit a violent act of terror. They also discuss honeypots and counter-propaganda techniques that can be used to rehabilitate radicalized users back to normal users. The disadvantage of keyword based flagging approach is the large number of false alarms. David and Morcelli \cite{decary2011gang} present a keyword based search to detect several criminal organizations and gangs on Twitter \& Facebook. They discuss a study of analyzing the presence of organized crime and how these gangs use social media platforms to recruit new members, broadcast their messages and coordinate their illegal activities on web $2.0$. They perform a qualitative analysis on $28$ groups and compare their organized crime between $2010$ and $2011$ on Facebook.\\
\begin{figure*}[t]
\centering{
\includegraphics[width=0.65\textwidth]{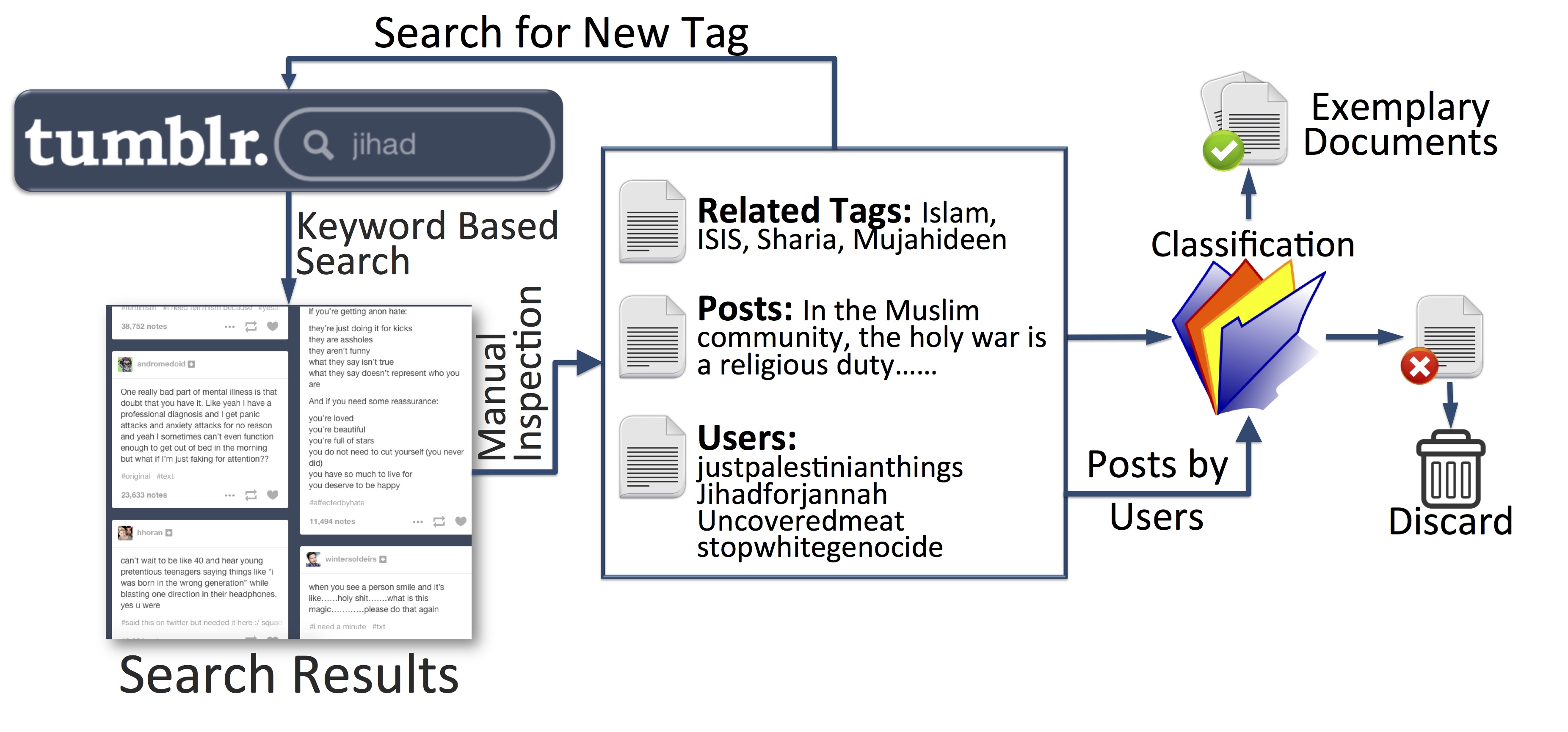}}
\caption{\label{exemplary_doc}Flow Sequence of Exemplary Data Collection Process}
\end{figure*}
\indent Agarwal et. al. propose a one-class classification model to identify hate and extremism promoting tweets \cite{sureka2014learning}. They conducted a case study on Jihad and identified several linguistic and stylistic features from free form text such as presence of war, religious, negative emotions and offensive terms. They conduct experiments on large real world dataset and demonstrate a correlation between hate promoting tweets and discriminatory features. They also perform a leave-p-out strategy to examine the influence of each feature on classification model.\\
\indent In context to existing work, the study presented in this paper makes the following unique contributions extending our previous work \cite{agarwal2015topical}:
\begin{enumerate}
\item We present an application of topical crawler based approach for locating extremism promoting bloggers on Tumblr. While there has been work done in the area of topical based crawling of social media platforms, to the best of our knowledge this paper is the first study on topical crawling for navigating connections between Tumblr bloggers.
\item We conduct experiments on large real world dataset to demonstrate the effectiveness of one class classifier and filtering hate promoting blog posts (text). We retrieve Tumblr blogger profiles and their links with other hate promoting bloggers and apply Social Network Analysis to locate strongly connected communities and core bloggers.
\end{enumerate}
\section{Experimental Setup}\label{experimental_setup}
\begin{figure}[t]
\centering{
\includegraphics[width=0.48\textwidth]{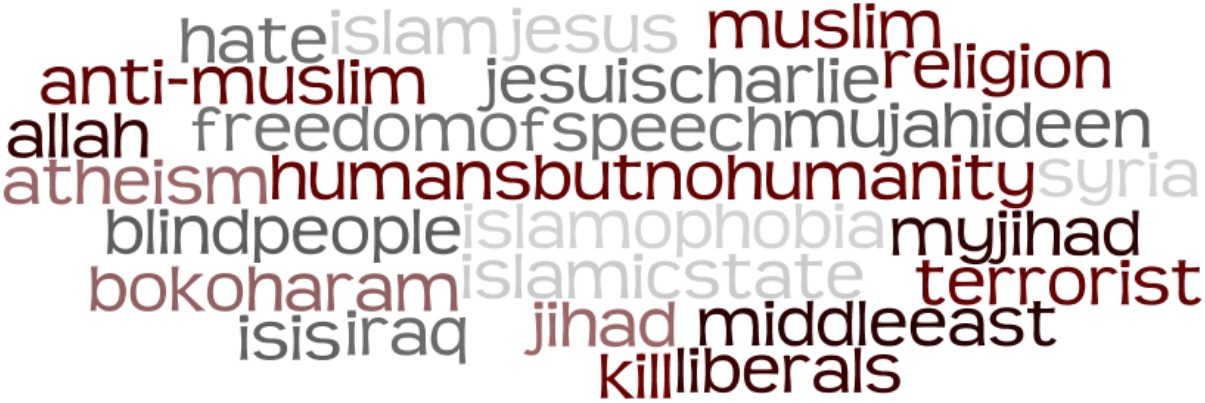}}
\caption{\label{cloud_terms}A Word Cloud of Key Terms Commonly Used by Extremist Bloggers}
\end{figure}

\begin{figure}[t]
\centering{
\includegraphics[width=0.45\textwidth]{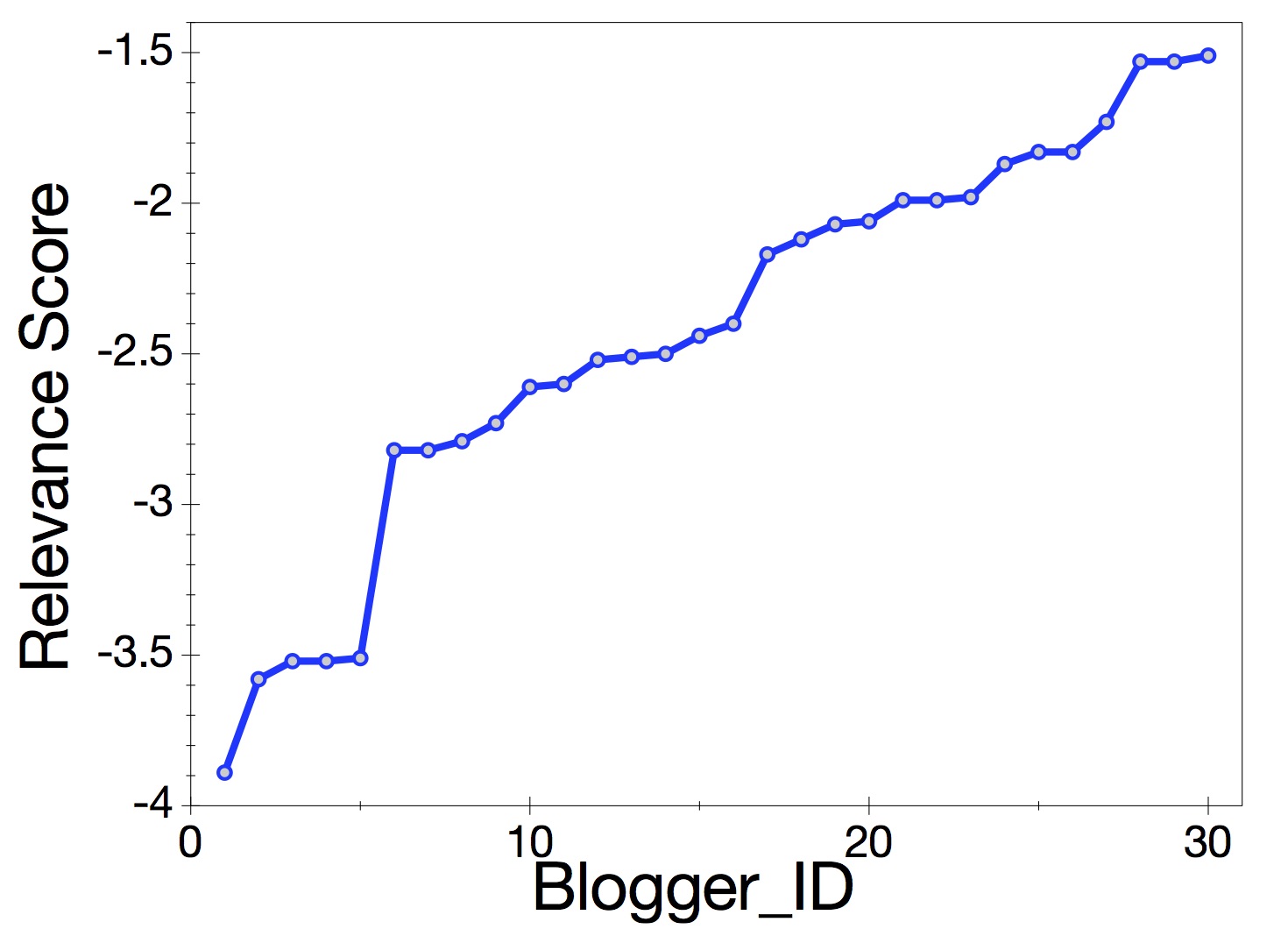}}
\caption{\label{relscore_threshold}Illustrating the Relevance Score Statistics of Positive Class Bloggers}
\end{figure}

We conduct our experiments on an open source and real time data extracted from Tumblr micro-blogging website. In a social networking website, a topical crawler extracts the external link to a profile and returns the nodes that are relevant to a defined topic. We define the relevance of a node based on extent of similarity of it's activity feeds and training document. Topical crawler learns the features and characteristics from these training documents and classify a profile to be relevant. Figure \ref{exemplary_doc} illustrates the general framework to obtain these documents. As shown in Figure \ref{exemplary_doc}, we implement a bootstrapping methodology to collect the training samples. We perform a manual search on Tumblr and create a lexicon of popular and commonly used tags associated with hate promoting posts. Figure \ref{cloud_terms} shows a word cloud of such terms. To collect our training samples, we perform a keyword (search tag) based flagging and extract their associated textual posts. We also acquire the related tags and the linked profiles (users who made these posts). We expand our list of keywords by extracting associated tags from these posts and their related tags. We run this framework iteratively until we get a reasonable number of exemplary documents ($400$ training samples). As mentioned above, we train our classifier for only hate promoting users. Therefore, the training documents contain the content and caption of only extremist posts.\\
\indent We use these linked bloggers and posts to compute the threshold value for language modeling. We take a sample of $30$ bloggers and compare their posts with the exemplary documents. For each blogger we get a relevance score. To compute the threshold value for similarity computation we take an average of these scores. Figure \ref{relscore_threshold} illustrates the relevance score statistics of each blogger (Sorted in increasing order). We notice that $80\%$ of the bloggers have relevance score between $-2.7$ and $-1.5$. We take average (turns out to be $-2.58$) of these scores to avoid the under-fitting and over-fitting of bloggers during classification. 
\begin{figure}[t]
\centering{
\includegraphics[width=0.48\textwidth]{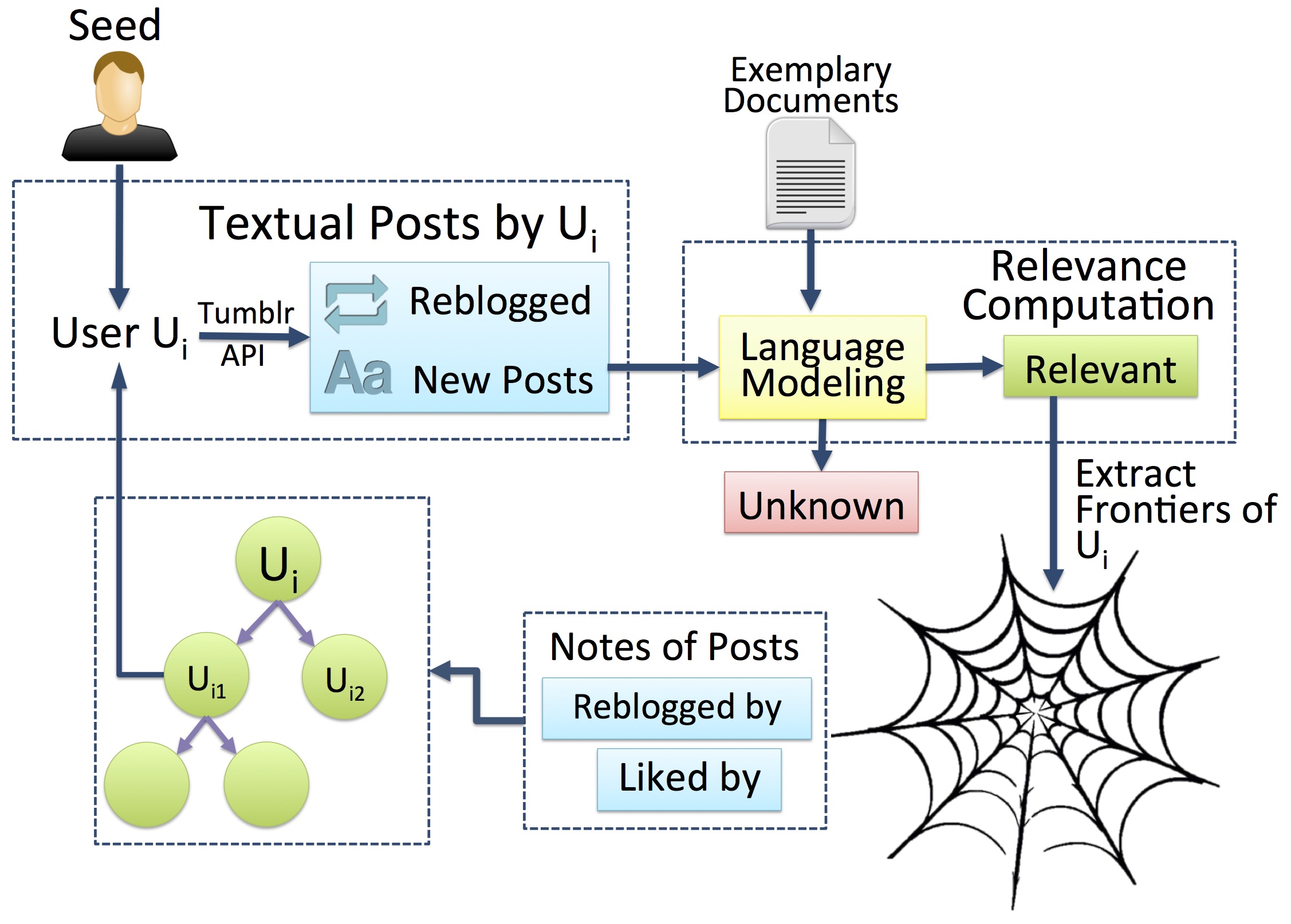}}
\caption{\label{framework}Proposed Architecture for Extremist Community Detection. Source: Agarwal et. al. \cite{agarwal2015topical}}
\end{figure}
\section{Research Methodology}\label{methodology}
In this section, we present the general research framework and methodology of proposed approach for classifying extremist bloggers on Tumblr (refer to Figure \ref{framework}). The proposed approach is an iterative multi-step that uses a hate promoting blogger as a seed channel and results a connected graph where nodes represents the extremist bloggers and links represents the relation between two bloggers (like and re-blog). As shown in Figure \ref{framework}, proposed framework is a multi-step process primarily consists of four phases: i) extraction of activity feeds of a blogger, ii) training a text classification model and filtering hate promoting and unknown bloggers, iii) navigating through external links to bloggers and extracting linked frontiers and iv) traversing through spider network for selecting next blogger. In Phase $1$, we use a positive blogger \textit{$U_{i}$} (annotated as hate promoting during manual inspection) called as 'seed'. We extract $n$ number of textual posts (either re-blogged or newly posted by user)of \textit{$U_{i}$} by using Tumblr API \footnote{https://www.tumblr.com/docs/en/api/v2}. We further use Jsoup Java library \footnote{http://jsoup.org/apidocs/} to extract the content and caption of these posts. Tumblr allows users to post content in multiple languages. However, our focus of this paper is to mine only English language posts. Therefore, we perform data-preprocessing on all extracted posts and by using Java language detection library \footnote{https://code.google.com/p/language-detection/}, we filter all non-English language posts. In Phase $2$, we train our classification model over training samples (refer to Section \ref{experimental_setup}). We perform character level n-gram language modeling\footnote{http://alias-i.com/lingpipe/index.html} on English language posts and compute their extent of similarity against training samples. We classify a blogger as hate promoting based on the relevance score and computed threshold value (refer to Section \ref{experimental_setup}).\\
\indent If a channel is classified as hate promoting or relevant, we further proceed to Phase $3$ and extract the notes information for each posts (collected in phase $1$). Notes in a Tumblr post contains the information about bloggers who liked or re-blogged a post. These user hits on a post indicates the similar interest among bloggers who may or may not be direct followers of each other. We extract the Tumblr ids of profiles from notes for the following reasons: i) due to privacy policy, Tumblr does not allow users/developers to extract the followers list unless the list is public and ii) Tumblr facilitates users to track any number of search tags or keywords. Whenever a new post is published on Tumblr containing any of these tags, it appears on the dashboard of user and a blogger no longer need to follow the original poster. We manage a queue of all extracted bloggers and traverse through the network using Random Walk algorithm. We use uniform distribution to select next blogger and extract it's frontiers. We extract these bloggers until the graph converges without re-visiting a blogger. The proposed framework results into a connected graph that represents a Tumblr network. We perform social network analysis on the output graph to locate hidden virtual groups and extremist bloggers playing major roles in community.
\begin{algorithm}[t]
\SetAlgoLined\DontPrintSemicolon
\KwData{User $U$, Consumer Key $C_{k}$, Consumer Secret $C_{s}$, Search Tag $tag\_name$}
\KwResult{Text based posts made by User $U$ or associated with tag $tag\_name$}
\SetKwFunction{ExtractPost}{ExtractPost}
\SetKwFunction{SetParameters}{SetParameters}
\SetKwFunction{TaggedPost}{TaggedPost}
\SetKwFunction{BloggerPost}{BloggerPost}
\nl \SetParameters{}\;
\nl \TaggedPost{}\;
\nl \BloggerPost{}\;
Generate URL of post to fetch post content and caption\;
\nl \For {$\textbf{all}$ $post P \in Posts$}{
\nl Slug=P.getSlug()\;
\nl id=P.getID()\;
\nl URL="http://blog\_name.tumblr.com/post/id/slug"\;
\nl Document=Jsoup.connect(URL).get()\;
\nl post\_content=Document.getDescription()\;
\nl post\_caption=Document.getTitle()\;
}
SetParameters() \{\;
\nl Authenticate the client via API Keys $C_{k}$ and $C_{s}$\;
\nl params.put("type", "text")\;
\nl params.put("filter", "text")\;
\nl params.put("reblog info", true)\;
\nl params.put("notes info", true)
\}\;
TaggedPost() \{\;
\nl Posts = client.tagged($tag\_name$, params)
\}\;
BloggerPost()\{\;
\nl Posts = client.tagged($tag\_name$, params)
\}\;
\captionsetup{justification=centering}
\caption{Extracting Textual Posts on Tumblr}\label{get_post}
\end{algorithm} 
\section{Solution Implementation} 
A topical crawler starts from a seed node, traverses in a graph navigating through some links and returns all relevant nodes to a given topic. In proposed solution approach we divide our problem into three sub-problems. First we classify the given seed node $S$ as hate promoting or unknown according to the published post (originally posted by blogger or re-blogged from other Tumblr users). Second, if the node is relevant then we extend this node into it's frontiers and it further leads us to more hate promoting bloggers. In third sub-problem, we perform topical crawling on Tumblr network and use random walk algorithm to traverse along the graph.
\subsection{Retrieval of Published Posts}
Algorithm \ref{get_post} describes the method to search Tumblr posts using keyword based flagging and extraction of posts published by a given blogger. The work presented in this paper focus on mining textual metadata on Tumblr therefore we set a few parameters and extract only text based posts for further analysis. For each blogger we set the limit of $100$ posts published recently. Function \SetParameters{} (steps $12$ and $13$) filters the search results and displays only the textual posts (quote, chat, text, url). Function \TaggedPost{} with given parameters search for text posts that exclusively contain given tag name. \BloggerPost{} fetches the textual posts published by given blogger ID. Both the functions make a Tumblr API request to fetch these data. Function \ExtractPost{} filters the response and extract body content \& caption of each post. Tumblr API allows us to only extract the summary of large posts. Therefore we use HTML parsing for extracting the whole message in blog post. In steps $4$ to $7$, we generate the URL from post summary and id to fetch the remaining post details. ID is a unique identifier of Tumblr posts and slug is a short text summary of that post which is appended in the end of every URL. We invoke this URL using Jsoup library and parse the HTML document to get the post content.
\begin{algorithm}
\SetAlgoLined\DontPrintSemicolon
\KwData{Blogger $U$}
\KwResult{Frontiers $F<name, type>$ of $U$}
\SetKwFunction{ExtractFrontiers}{ExtractFrontiers}
ExtractFrontiers{$U$}\{\;
\nl \SetParameters{}\;
\nl Posts=\BloggerPost{$U$}\;
\nl \For {$\textbf{all}$ $post P \in Posts$}{
\nl Notes=P.getNotes()\;
\nl \For {$\textbf{all}$ $Note N \in Notes$}{
\nl Linked\_Blog\_Name=N.getBlogName()\;
\nl Note\_Type=N.getType() \qquad Liked or Reblog\;
\nl Frontiers $F$.add(Linked\_Blog\_Name, Note\_Type)\;
}
}
\}\;
\captionsetup{justification=centering}
\caption{Extracting Frontiers of a Given Blog}\label{get_frontiers}
\end{algorithm}
\begin{figure}[t]
\centering{
\includegraphics[width=0.48\textwidth]{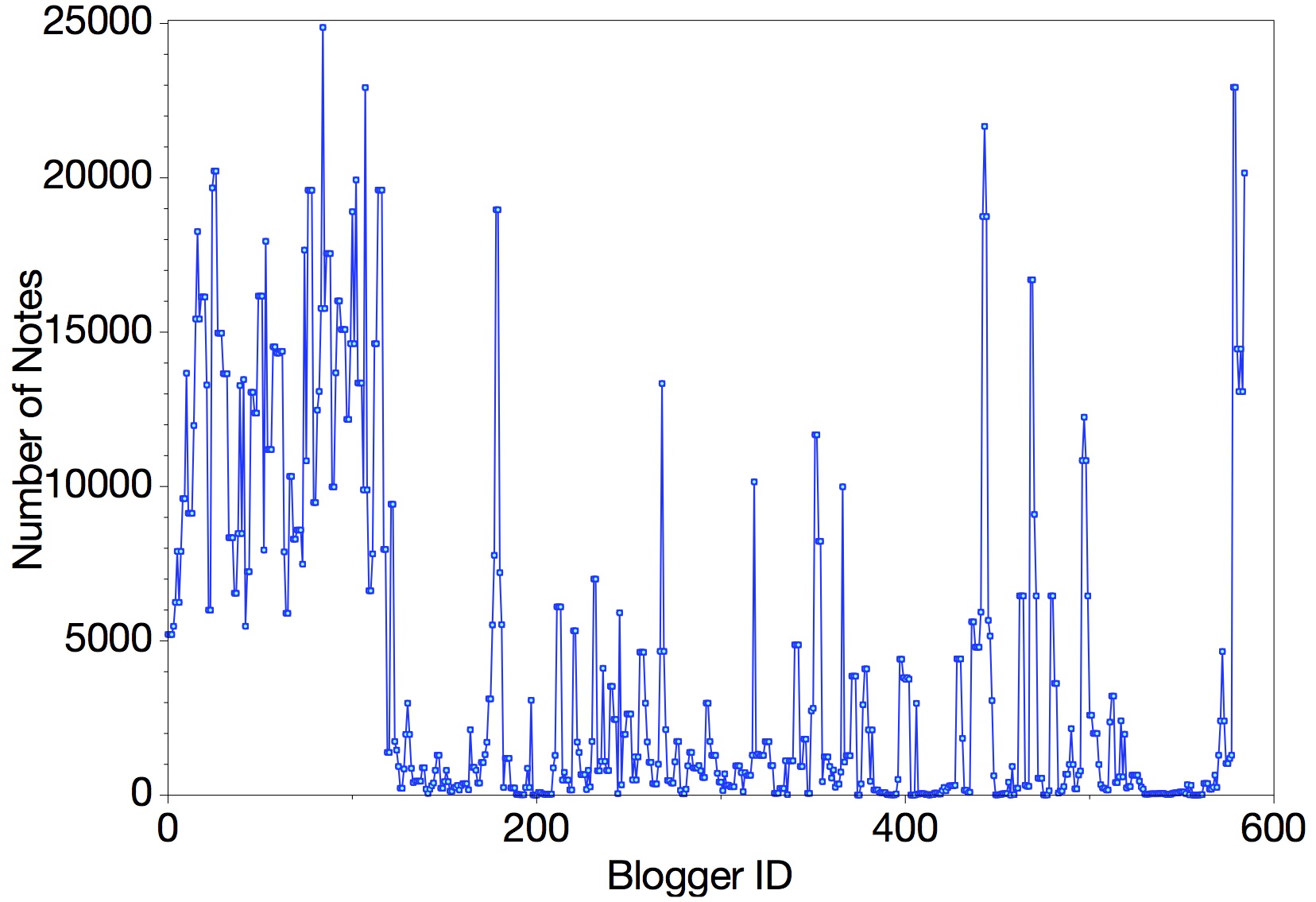}}
\caption{\label{notes}Illustrating the Number of Notes For Each Blogger Traversed in Topical Crawler}
\end{figure}
\begin{algorithm}[t]
\SetAlgoLined\DontPrintSemicolon
\KwData{$S$, $th$, $N_{g}$, $S_{g}$, $W_{g}$, $D_{e}$}
\KwResult{Directed Graph $G$}
\nl \SetParameters{}\;
\nl $U_{i}=S$, F.add($S$)\;
\SetKwFunction{TopicalCrawler}{TopicalCrawler}
TopicalCrawler{$S$}\{\;
\nl \While{($graph size < S_{g}$ OR $F.size>0$)}{
\nl Posts=\ExtractPost{$U_{i}$}\;
\nl Relevance\_Score = LanguageModeling($D_{e}$, $Posts$, $N_{g}$)\;
\nl \eIf{($score > $th)}{
\nl Linked\_Users=\ExtractFrontiers{$U_{i}$}\;
\nl ProcessedNodes PN.add($U_{i}$)\;
\nl \For {$\textbf{all}$ $LU \in Linked\_Users$}{
\nl \eIf{(!(F.contains($LU$) AND (PN.contains($LU$)))}{
\nl F.add($LU$)\;}{
\nl Discard the node $LU$\;}
}}{
\nl Discard the node $U_{i}$\;}
\nl Compute the Markov Chain over graph $G$\;
\nl New\_Blogger= node with maximum probability in Markov chain array\;
\nl F.remove(New\_Blogger)\;
\nl \TopicalCrawler{New\_Blogger}\;
}
\}\;
\captionsetup{justification=centering}
\caption{Graph Traversal Using Random Walk Algorithm}\label{algo_random_walk}
\end{algorithm}
\subsection{Retrieval of External Links to Bloggers} Algorithm \ref{get_frontiers} describes the steps to extract frontiers of a given node $U$. Due to the privacy policies, Tumblr API does not allow developers to extract subscriptions and followers of a Tumblr user. The link between two bloggers indicates the similar interest between them so that number of frontiers vary for every post published by a blogger. For each user, we extract $25$ bloggers for each relation i.e. users who have liked and re-blogged that post recently. If a blogger $B_{1}$ re-blog and as well as likes a post published by another blogger $B_{2}$ then in the graph $G$, we create an edge with both labels i.e. ($B_{1}$, $B_{2}$, <like, re-blog>). To avoid the redundancy we extract one more frontier who have either liked or re-blogged the post recently. To extract the linked bloggers of a Tumblr user we first need to extract the posts made by $U$. We can extract notes information only when notes and re-blog information parameters are set to be true (refer to steps $14$ and $15$ of Algorithm \ref{get_post}). As described in Algorithm \ref{get_frontiers}, in step $4$, we extract notes for each textual post (hate promoting) made by User $U$. In steps $5$ to $8$, we extract the name of unique bloggers who liked and re-blogged the post $P$. In step $8$, $F$ represents the list of frontiers and relation of $U$ with each frontier. We maintain a list of all processed bloggers and the number of hit counts on their recent $100$ posts. These number of notes varied from $0$ to $25$K therefore we perform smoothing on data points and plot median of these values. Figure \ref{notes} shows the statistics of number of notes collected on $100$ posts of each blogger extracted during topical crawler. Figure \ref{notes} reveals that overall number of hit counts (number of reblog and like) for extremism promoting users is very high. These hit counts reveals the popularity of extremist content and the number of viewers connected to such bloggers. 
\subsection{Topical Crawler Using Random Walk}
Algorithm \ref{algo_random_walk} describes the proposed crawler for locating a group of hidden extremist bloggers on Tumblr. The goal of this algorithm is to compare each blogger against training examples and then connecting all positive class (hate promoting or relevant) bloggers. Algorithm \ref{algo_random_walk} takes several inputs: seed blogger (positive class user) $S$, size of the graph $S_{g}$ i.e. maximum number of nodes in a graph, width of the graph $W_{g}$ i.e. the maximum number of frontiers or adjacency nodes for each blogger, a set of exemplary documents $D_{e}$, threshold $th$ and n-gram value $N_{g}$ for relevance computation. We create a list of $30$ positive class bloggers extracted during experimental setup (refer to section \ref{experimental_setup}) and compute their relevance score against the exemplary documents. We take an average of these scores and compute the threshold value for language modeling. We use n-gram language modeling ($N_{g}$=$3$) to build our statistical model. Algorithm \ref{algo_random_walk} is a recursive process that results into a cyclic directed graph. We run this algorithm until we get a graph of size $S$ ($1000$ bloggers) or there is no node left in the queue for further extension. We perform a self-avoiding random walk that means we make sure a node is never being re-visited. If a node re-appears in the frontiers list then there are two possibilities: 1) the frontier has already been processed (extended or discarded based upon the relevance score- Steps $4$ to $7$). If it exists in the processed nodes list then we create a directed edge between the node and it's parent and avoid further extension. 2) If the re-appearing node is in frontiers list and is not yet processed, we created a directed edge in the graph and continue the traversal.\\
\indent The topical crawler is a recursive process that adds and removes nodes after each iteration. The resultant graph is dynamic and not irreducible that means given a graph $G(V, E)$, if there is a directed edge between two nodes $u$ and $v$, it is not necessary that there exists a directed path from $v$ to $u$. Consider that object (topical crawler) processed node $i$ at time $t-1$. In the next iteration object moves to an adjacency node of $i$. The probability that object moves to node $j$ at time $t$ is $\frac{1}{d^{+}_{i}}$ when there exists a direct edge from $i$ to $j$. $M_{ij}$=$\frac{1}{d^{+}_{i}}$ denotes the probability to reach from $i$ to $j$ in one step where $d^{+}_{i}$ is the out-degree of node $i$. Therefore we can define:\\
\begin{equation}\label{probability}
M_{i,j}=
\left\{
\begin{aligned}
\frac{1}{d^{+}_{i}},\ \ \ if \ (i,j) \ is \ an \ edge \ in \ digraph \ G
\\
0,\ \ \ \ \ \ \ \ \ \ \ \ \ \ \ \ \ \ \ \ \ \ \ \ \ \ \ \ \ \ \ \ \ \ \ otherwise
\end{aligned}
\right\}
\end{equation}

Therefore for each vertex $i$, the sum of the probability to traverse an adjacency node of $i$ is $1$.\\
\begin{equation}\label{Mij}
\forall i \sum_{j \in A(i)} M_{ij}=1\\
\end{equation}
Where, A(i) denotes the list of adjacency nodes $i$. In random walk on graph $G$ topical crawler traverse along the nodes according to the probability of $M_{ij}$. Graph $G$ is a dynamic social networking graph, therefore we compute a Markov chain M after each iteration and compute the probability matrix over graph $G$. Markov chain is a random process where the probability distribution of node $j$ depends on the current state of matrix. The probability matrix $M^{k}$ gives us a picture of graph $G$ after $k$ iterations of topical crawler. Using this matrix, we compute the probability distribution $P$ that object moves to a particular vertex. $P^{k}$ is the probability distribution of a node $j$ after $k$ iteration then probability of $i$ to be traversed in $k+1^{th}$ iteration is the following:
\begin{equation}\label{prob_dis}
P^{k+1}=P^{k}.M \\
where, P^{k}= P^{0}*M^{k}\\
\end{equation}

Where $P^{0}$ is the initial distribution fixed for the seed node.
\begin{table}[t]
\caption{Confusion Matrix and Accuracy Results for Unary Classification Performed During Topical Crawler. Source: Agarwal et. al. \cite{agarwal2015topical}}\label{results}
\centering
\subtable[\label{conf_matrix}Confusion Matrix]{%
\begin{tabular}{|c|c|c|c|}
\hline
\multicolumn{2}{|c|}{} &\multicolumn{2}{c|}{\textbf{ Predicted}}\\
\hline
\multicolumn{1}{|c|}{} & &\textbf{Positive }& \textbf{ Unknown}\\
\cline{2-4}
\multirow{2}{*}{\textbf{Actual} }& \multirow{1}{*}{ \textbf{Positive}} & 290 & 45\\
\cline{2-4}
& \multirow{1}{*}{ \textbf{ Unknown} } & 92 & 173\\
\hline
\end{tabular}}%
\qquad
\subtable[\label{acc_results}Accuracy Results]{%
\begin{tabular}{|c|c|c|c|}
\hline
\textbf{Precision}&\textbf{Recall}&\textbf{F-Score}&\textbf{Accuracy}\\
\hline
0.75&0.86&0.80& 0.77\\
\hline
\end{tabular}}%
\end{table}
\section{Experimental Results}
\subsection{Topical Crawler Results}
We hired $30$ graduate students from different departments as volunteers to annotate all $600$ bloggers processed or traversed during the topical crawling. We provided them guidelines for annotation and to remove the bias, we performed horizontal and vertical partitioning on the bloggers' dataset. We divided annotators into $10$ different groups, $3$ members each. We asked each group to to annotate $60$ bloggers. Therefore, we got $3$ reviews for each blogger. We used majority voting approach (a blogger labeled as X by at least two annotators) for annotating each blogger as hate promoting or unknown. We compute the effectiveness of our classification by using precision, recall and f-measure as accuracy matrices. Table \ref{results} shows the confusion matrix and accuracy results for unary classification performed during graph traversal. Table \ref{conf_matrix} reveals that among $600$ bloggers, our model classifies $382$($290$+$92$) bloggers as hate promoting and $218$($173$+$45$) bloggers as unknown with a misclassification of $13$\% and $34$\% in predicting extremist and unknown bloggers respectively. Table \ref{acc_results} shows the standard information retrieval matrices for accuracy results. our results shows that the both the precision and recall are high, as it is important to reduce the number of false alarms and also not to miss an extremist blogger in order to locate their communities. We use F-score as the accuracy metrics for our classifier and the results reveal that we are able to predict extremist blogger with an f-score of $80$\%.
\begin{table*}[ht!]
\captionsetup{justification=centering}
\caption{Illustrating The Network Level Measurements for Topical Crawler.Dia= Network Diameter, Mod= Modularity, ACC= Average Clustering Coefficient, IBC= In- Betweenness Centrality, CC= In- Closeness Centrality, \#SCC= Number of Strongly Connected Components.}\label{social_network_measurements}
\centering
\begin{tabular}{|c|c|c|c|c|c|c|c|c|}
\hline
\textbf{Graph}&\textbf{\#Nodes}&\textbf{\#Edges}&\textbf{Dia}&\textbf{\#SCC}&\textbf{\#ACC}&\textbf{\#Mod}&\textbf{IBC}&\textbf{ICC}\\
\hline
\textbf{TC}&382&275&4&137&0.026&12.00&11.36&0.20\\
\hline
\textbf{LB}&27&60&1&21&0.0231&1.307&0&0.38\\
\hline
\textbf{RB}&355&215&6&185&0.021&7.01&6.284&0.40\\
\hline
\end{tabular}
\end{table*}
\begin{figure*}[ht!]
\centering{
\subfigure[\label{graph_gephi}]{\includegraphics[width=0.32\textwidth]{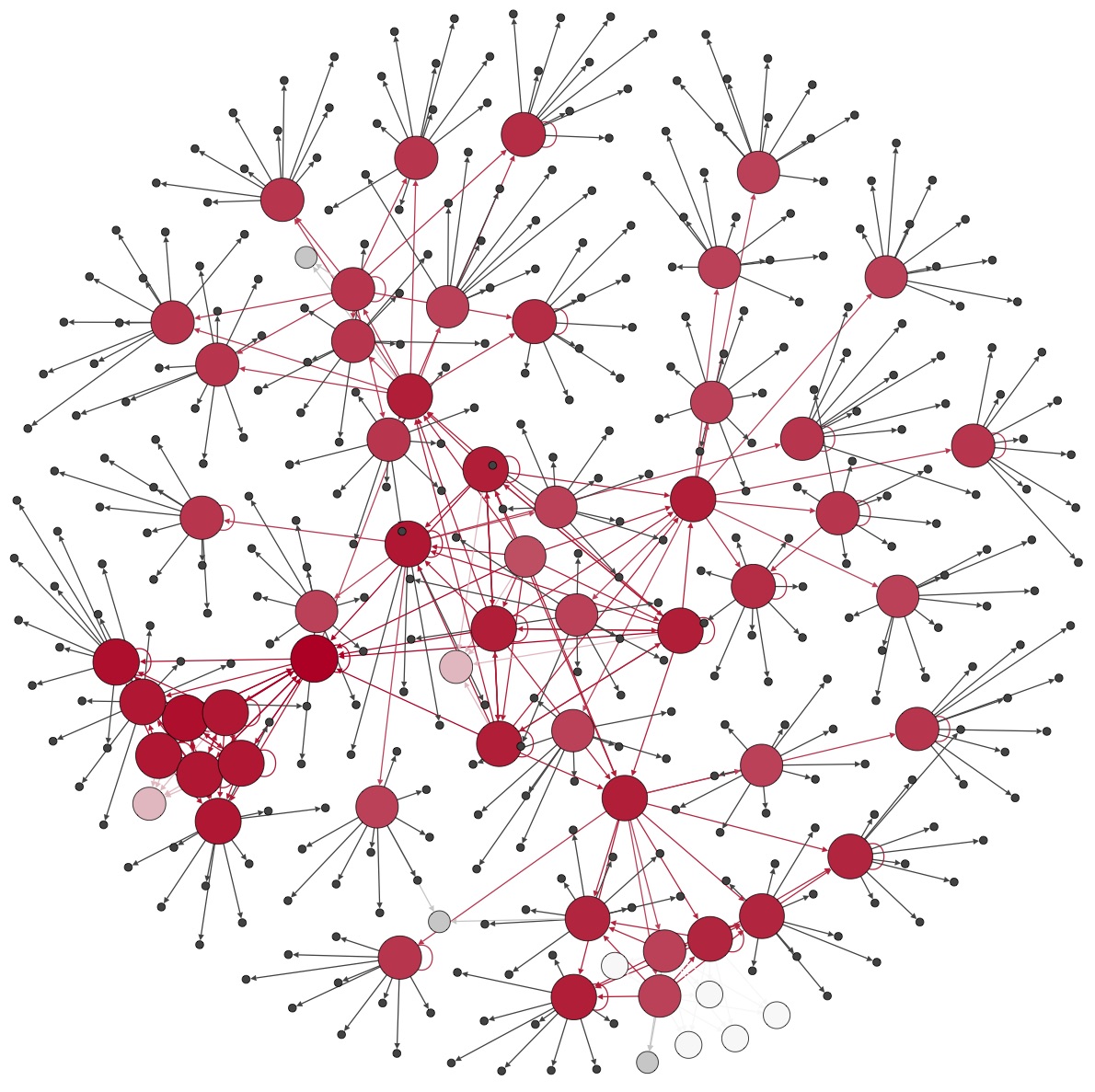}}
\subfigure[\label{graph_like}]{\includegraphics[width=0.32\textwidth]{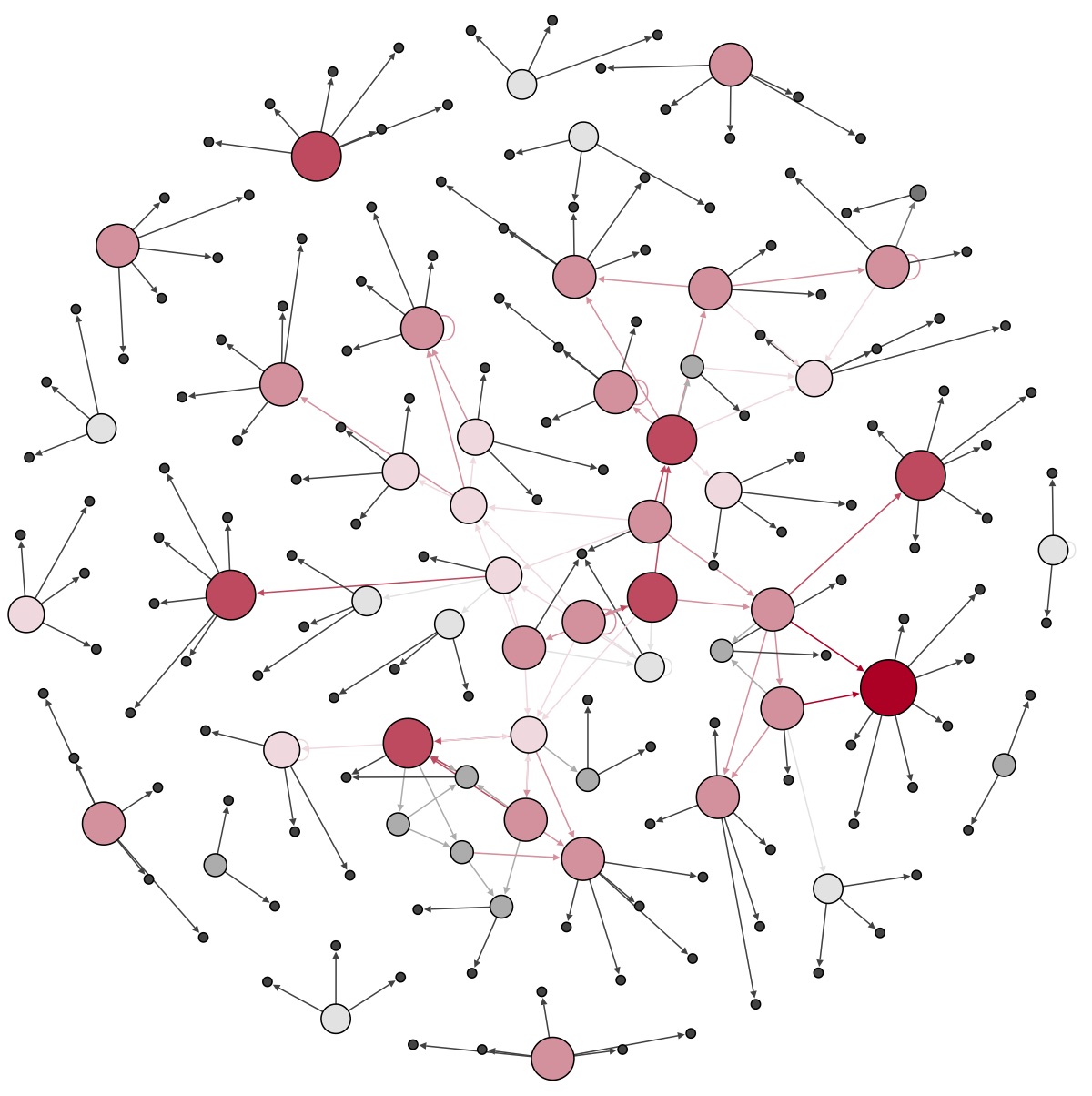}}
\subfigure[\label{graph_re-blog}]{\includegraphics[width=0.32\textwidth]{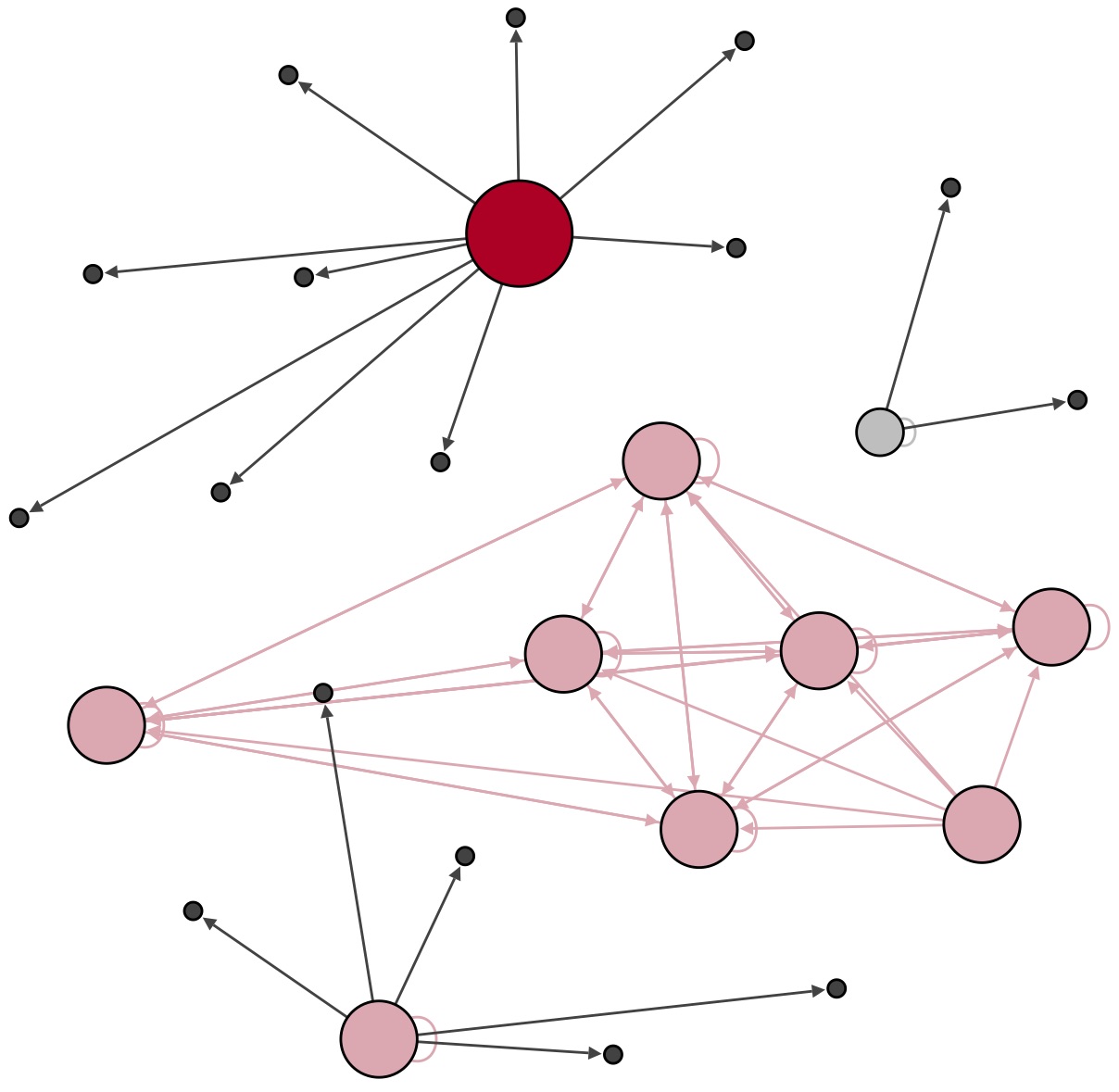}}}
\captionsetup{justification=centering}
\vspace{-0.2cm}
\caption{Cluster Representation of Social Network Graphs- Topical Crawler using Random Walk (a), 'Posts Liked by' (b) and 'Posts Re-blogged By' (c)}\label{community}
\end{figure*} 
\subsection{Social Network Analysis}
We perform social network analysis on topical crawler's network resulted into a directed graph $G(V, E)$, where $V$ represent a set of Tumblr bloggers accounts and $E$ represent a directed edge between two bloggers. We define this edge as a relation having two labels 'posts liked by' and 'post re-blogged by'. To examine the effectiveness of these relations we generate two independent networks exclusively for 'like' and 're-blog' links between bloggers. Figure \ref{community} illustrates the representations of these networks. In each graph, size of the node is directionally proportional to its out-degree. A node with maximum number of adjacency vertices is biggest in size. Colours in the graph represents the clusters of nodes having similar properties. Here, we define the similarity measure as the ratio of out-degree and in-degree.\\
\indent We also perform several network level measurements on these graphs. Table \ref{social_network_measurements} reveals that re-blogging is a strong indicator of links between two Tumblr profiles. Here, we observe that the graphs generated for topical crawler and re-blogging link have same pattern in network measurements. Both graphs are dense (also evident from the Figure \ref{community}) and have higher modularity in comparison to the graph created for 'liked' link. Table \ref{social_network_measurements} and Figure \ref{community} also reveal that by navigating through re-blogging links we can locate large number of connected components in a extreme right community. While following 'like' as a link we are able to detect small number of connected blogs. Though as illustrated in Figure \ref{graph_like}, we can not completely avoid this feature as a set of blogs extracted using this link are irreducible. Table \ref{social_network_measurements} also shows that the graph created for 'like' relation has slightly larger value for average clustering co-efficient. This is because the number of nodes in the graph is very less and a major set of these nodes is strongly connected. Higher value of In-between centrality shows the presence of bloggers who are being watched by a large number of users. As the Figure \ref{graph_re-blog} shows there are many users which are not directly connected to each other (shown in red colour) but has a huge network of common bloggers. These disjoint bloggers are two or three hop away and are connected via other bloggers (having second largest number of adjacency nodes). These bloggers are connected with maximum number of other bloggers present in the graph and has a wide spread network in extreme right communities. These nodes have the maximum closeness centrality and play central role in the community. Nodes represented as black dots have minimum number of out-degree. They don't have a directed path to the central users or original source of extremist posts. Based upon our study we find these bloggers to be the target audiences who share these posts in their own network. These users are very crucial for such communities though they don't play a major role in the network.
\section{Conclusions}
In this paper, we perform a case study on Jihadist groups and locate their existing extreme right communities on Tumblr. We conduct experiments on real world dataset and use topical crawler based approach to collect textual data (published posts) from Tumblr users. We perform one class classification and identify hate promoting bloggers according to the content present in their posts. We use random walk algorithm for graph traversal and extract exclusive links to these bloggers. We conclude that by performing social networking analysis on a graph (vertices are the Tumblr bloggers and edges are the links among these bloggers: re-blog and like) we are able to uncover hidden virtual communities of extremist bloggers with an accuracy of $77$\%. We compute various centrality measures to locate the influential bloggers playing major roles in extremist groups. We also investigate the effectiveness of link features (likes and re-blogs) in order to find the communities. Our results reveal re-blogging is a strong indicator and a discriminatory feature to mine strongly connected communities on Tumblr.\\
\indent We perform a manual inspection on Tumblr and perform a characterization on several hate promoting posts. Our study reveals that these posts are very much popular among extremist bloggers and get large number of hits. These posts are published targeting some specific audiences. Keywords present in the blog content, tags associated with post and comments by other bloggers are clear evidence of hate promotion among their viewers.
\balance

\end{document}